\documentclass[conference]{IEEEtran}

\usepackage{graphicx}%
\usepackage{bm}%
\usepackage{amsfonts}
\usepackage{amssymb}
\usepackage{url}
\usepackage{times}
\usepackage{subfigure}
\usepackage{enumitem}
\usepackage{amsmath} 
\usepackage{latexsym}
\usepackage{hhline}
\usepackage{cite}
\usepackage{caption}
\usepackage{algorithm}
\usepackage{algpseudocode}
\usepackage{multirow} 
\usepackage{xcolor}
\usepackage{epstopdf}
\usepackage{aligned-overset}

\allowdisplaybreaks

\begin{document}

\bstctlcite{bibfbcs:BSTcontrol}

\title{Channel Estimation for OFDM via Delay–Doppler Refinement}
\author{\IEEEauthorblockN{
Mingcheng Nie\IEEEauthorrefmark{1}, 
Hao Chang\IEEEauthorrefmark{1},
Xiaoqi Zhang\IEEEauthorrefmark{2},
Junkai Liu\IEEEauthorrefmark{3},  
Wibowo Hardjawana\IEEEauthorrefmark{1},
Branka Vucetic\IEEEauthorrefmark{1},
Yonghui Li\IEEEauthorrefmark{1}
}
\IEEEauthorblockA{
\IEEEauthorrefmark{1}The University of Sydney, Sydney, Australia\\
\IEEEauthorrefmark{2}University of Technology Sydney, Sydney, Australia\\
\IEEEauthorrefmark{3}Fudan University, Shanghai, China\\
\emph{(Invited Paper)}
}}

\markboth{Journal of \LaTeX\ Class Files,~Vol.~14, No.~8, August~2021}%
{Shell \MakeLowercase{\textit{et al.}}: A Sample Article Using IEEEtran.cls for IEEE Journals}


\maketitle

\begin{abstract}

In this paper, we propose a novel channel estimation (CE) algorithm for orthogonal frequency division multiplexing (OFDM) systems that exploits the unique characteristics of the delay-Doppler (DD) domain channel. Specifically, the time-frequency (TF) domain input-output relationship (IOR) is derived in a compact form by focusing solely on the non-zero elements of the TF domain channel matrix. Based on this compact IOR, a coarse TF domain CE is first performed using a linear minimum mean square error estimator. Then, the resultant TF domain estimates are transformed to the DD domain through a unitary transformation for further refinement.
We reveal that the effective DD domain channel matrix can be viewed as an aggregation of multiple DD domain channel responses with different phase shifts. This allows us to devise a threshold-based estimation for DD domain channel parameters with high accuracy. The estimated DD domain channel parameters are then applied to form a refined estimate of TF domain channel. Our numerical results demonstrate that the proposed method can achieve substantial performance gains over conventional OFDM channel estimation techniques under the same pilot deployment.

\end{abstract}

\section{Introduction}

Orthogonal frequency division multiplexing (OFDM) has been widely adopted in modern wireless networks due to its robustness against frequency-selective fading and simple implementation~\cite{liu2024fast}.
However, the success of OFDM relies on strict subcarrier orthogonality, which becomes unrealistic in high-mobility scenarios~\cite{11373535,xiaoqi2025otfs}. In such cases, high Doppler effects introduce non-uniform frequency shifts across subcarriers, breaking the orthogonality among the subcarriers and resulting in severe inter-carrier interference (ICI). Furthermore, reduced channel coherence time and limited pilot resources for high spectral efficiency lead to significant performance loss in conventional time-frequency (TF) domain estimation methods~\cite{nie2024uplink,xiaoqiprior2025}.
Recently, delay-Doppler (DD) domain channel modeling and related waveforms, such as orthogonal time-frequency space (OTFS)~\cite{hadani2017orthogonal}, have attracted significant attention for their ability to enable reliable communication over high-mobility channels. By transforming the rapidly time-varying TF channel into a quasi-static, sparse DD representation, DD communication offers a new perspective for channel estimation (CE). For instance, a DD domain CE scheme based on an embedded pilot~\cite{raviteja2019embedded} achieved accurate estimates by leveraging the quasi-static nature of the DD channel response.

However, DD domain channel estimations rely on specific DD domain pilot arrangements~\cite{nie2023improving}, which are not compatible with practical OFDM frame structures. For example, the commonly applied DD domain embedded pilot~\cite{raviteja2019embedded} spreads across the whole TF domain, introducing considerable interference to the information symbols in OFDM systems. 
This naturally raises a fundamental question: is it possible to exploit the advantages of DD domain channels while still employing OFDM pilots transmitted in the TF domain? While providing an effective solution to this question could dramatically enhance conventional OFDM performance, only limited studies have explored this direction. For example, \cite{hu2024cross} proposed a cross-domain channel estimation (CDCE) approach for OTFS, where TF channel estimation is first performed via single-tap equalization (STE) and then refined in the DD domain using a cross-correlation function. However, \cite{hu2024cross} assumes an ICI-free TF domain channel, leading to model mismatch in practice and thereby a severe error floor. In contrast, our previous work~\cite{nie2025novel} proposed a novel CDCE for OFDM systems, which significantly improved estimation performance without exhibiting an error floor caused by ICI. The proposed method first transforms TF domain pilot sequences and received signals into the DD domain for coarse estimation via two-dimensional (2D) twisted-convolution. Then, the TF domain estimation is formulated as a least-squares problem and solved either through sparse recovery or a pseudo-inverse, depending on the dictionary matrix dimensions.

In this paper, we propose an alternative CE method for OFDM transmissions based on DD domain refinement in the presence of high channel Doppler. The proposed method first performs a coarse TF domain channel estimation using a linear minimum mean square error (LMMSE) estimator, after which the estimates are transformed to the DD domain for further refinement. 
Particularly, we reveal that the effective DD domain
channel matrix can be viewed as an aggregation of multiple
DD domain channel responses with different phase shifts in the presence of integer channel delay and Doppler. Based on this, a threshold-based DD channel parameter estimation is developed and applied to each column of the effective DD domain
channel matrix. The estimated parameters from each matrix column are then combined and used for TF domain channel matrix construction with high accuracy. Our numerical results demonstrate that the proposed method significantly improves channel estimation performance, in terms of normalized mean square error (NMSE), compared with conventional OFDM channel estimation schemes.

\emph{Notation:}
$(\cdot)^{\rm{H}}$, $(\cdot)^{*}$ and $(\cdot)^{\rm{T}}$ denote the Hermitian, conjugate, and transpose, respectively; 
${\rm Tr}\{\cdot\}$ and ${\rm{vec}}\left( \cdot \right)$ denote the trace and the vectorization of a matrix; $[\cdot]_N$ represents the modulo operation w.r.t. $N$;
${{\bf{F}}_N}$ denotes the normalized discrete Fourier transform (DFT) matrix of size $N\times N$;
${\bf I}_M$ represents the $M\times M$ identity matrix;
``$ \otimes $" denotes the Kronecker product operator. 

\section{System Model}

\subsection{OFDM Transmissions over Doubly-Selective Channel}
We consider an OFDM system comprising $M'$ subcarriers with subcarrier spacing $\Delta f$, and $N$ OFDM symbols, each with duration $T$. The system operates under the critical sampling condition such that $\Delta f=\frac{1}{T} $. The transmitted symbols are represented in the TF domain by $\mathbf{X}'_\mathrm{TF}\in\mathbb{C}^{M'\times N}$. The time domain transmit signal $\mathbf{S}'$ is then obtained by applying the inverse DFT (IDFT) along the frequency dimension as
\begin{align}
    \mathbf{S}'=\mathbf{F}_{M'}^{\mathrm{H}}\mathbf{X}'_\mathrm{TF}.
\end{align}
By appending the CP of length $L_{\rm CP}$ to each OFDM symbol, we have
\begin{align}
    \mathbf{S}=\mathbf{A}_{\mathrm{CP}}\mathbf{S}'=\mathbf{A}_{\mathrm{CP}}\mathbf{F}_{M'}^{\rm H}\mathbf{X}'_\mathrm{TF},
\end{align}
which can be written in vector form as
\begin{align}
    \mathbf{s}=(\mathbf{I}_N\otimes\mathbf{A}_{\mathrm{CP}})\mathbf{s}'=(\mathbf{I}_N\otimes\mathbf{A}_{\mathrm{CP}}\mathbf{F}_{M'}^{\rm H})\mathbf{x}'_\mathrm{TF},
\end{align}
where $\mathbf{s}\in\mathbb{C}^{MN\times 1}$ denotes the time domain transmit signal after inserting OFDM CP and $M=M'+L_{\rm CP}$. Moreover, $\mathbf{A}_{\mathrm{CP}}=[\mathbf{G}_{\mathrm{CP}},\mathbf{I}_{M'}]^{\mathrm{T}}\in\mathbb{R}^{M\times M'}$ is the OFDM CP addition matrix, where $\mathbf{G}_{\mathrm{CP}}$ of size $M'\times L_{\mathrm{CP}}$ includes the last $L_{\mathrm{CP}}$ columns of the identity matrix $\mathbf{I}_{M'}$. 
On top of these operations, we apply an additional CP to the OFDM symbol sequence $\bf s$, to ensure that our derivation is consistent with the underlying transformations of OTFS.
Let $\tilde{\mathbf{A}}_{\mathrm{CP}}=[\tilde{\mathbf{G}}_{\mathrm{CP}},\mathbf{I}_{MN}]^{\mathrm{T}}\in\mathbb{R}^{(MN+L_{\mathrm{CP}})\times MN}$  be the OTFS CP addition matrix, where $\tilde{\mathbf{G}}_{\mathrm{CP}}$ of size $MN\times L_{\mathrm{CP}}$ has a similar structure to $\mathbf{G}_{\mathrm{CP}}$. Thus, we obtain
\begin{align}
\tilde{\mathbf{s}}=\tilde{\mathbf{A}}_{\mathrm{CP}}\mathbf{s}=\tilde{\mathbf{A}}_{\mathrm{CP}}(\mathbf{I}_N\otimes\mathbf{A}_{\mathrm{CP}})\mathbf{s}'.
\end{align}
Subsequently, the continuous time domain transmit signal can be obtained by applying a pulse shaping $p(t)$, yielding
\begin{align}
    s(t)=\sum\nolimits_{n=-L_{\mathrm{CP}}}^{MN-1}\tilde{{s}}[n] p(t-nT_{\mathrm{s}}),
\end{align}
where $\tilde{{s}}[n]$ denotes the $n$-th element of $\tilde{\mathbf{s}}$ and $T_{\mathrm{s}}$ is the time domain sampling period satisfying $T_s=\frac{T}{M'}$. Here, we focus on a time-varying channel, whose sparse representation in the DD domain is given by $h(\tau,\nu)=\sum\nolimits_{p=1}^{P}h_p\delta(\tau-\tau_p)\delta(\nu-\nu_p),$
where $h_{p}$, $\tau_{p}=\left(l_{p}+\imath_{p}\right)\frac{1}{M'\Delta f}$, and $\nu_{p}=\left(k_{p}+\kappa_{p}\right)\frac{1}{NT}$ denote the channel coefficient, delay, and Doppler shifts of the $p$-th path, respectively. Here, $l_{p}$ and $k_{p}$ represent the integer delay and integer Doppler indices, respectively, while $-{1}/{2}\le\imath_p\le {1}/{2}$ and $-{1}/{2}\le\kappa_p\le {1}/{2}$ represent the fractional delay and fractional Doppler indices, respectively. 
For simplicity, we assume that the transmission has sufficient delay and Doppler resolutions, i.e., the fractional delay and Doppler indices are zero.

After experiencing the time-varying channel, the continuous-time domain receive signal $r(t)$ can be expressed as
\begin{align}
    r(t)=\sum\nolimits_{p=1}^{P} h_p s(t-\tau_p)e^{j2\pi\nu_p(t-\tau_p)}+w(t),
\end{align}
where $w(t)$ is the complex additive white Gaussian noise (AWGN) process with zero mean and variance $N_0$. After matched-filtering, we obtain its discrete-time representation $\tilde{r}[m]$, whose $m$-th element is expressed as
\begin{align}
     \tilde{r}[m]=\sum\nolimits_{n=-L_{\mathrm{CP}}}^{MN-1}\tilde{{s}}[n]  g[m,n]+w[m].\label{g_first}
\end{align}
Here, we define the effective time domain channel as 
\begin{align}
    g[m,n] \triangleq\sum\nolimits_{p=1}^{P} h_p e^{j2\pi n \nu_p T_{\mathrm{s}}} A^*((n-m)T_{\mathrm{s}}+\tau_p,\nu_p),\label{g_element}
\end{align}
where $A(\tau_p,\nu_p)$ denotes the ambiguity function of pulse $p(t)$ with respect to delay $\tau_p$ and Doppler shift $\nu_p$, whose definition is given by
\begin{align}
    A(\tau_p,\nu_p)&\triangleq\int_{-\infty}^{\infty} p(t) p^*(t-\tau_p) e^{-j2\pi\nu_p(t-\tau_p)} dt.
\end{align}
By stacking all received symbols into a vector and removing the OTFS CP, we obtain the following compact representation of the received signal:
\begin{align}
    \mathbf{r}&=\tilde{\mathbf{R}}_{\mathrm{CP}}\tilde{\mathbf{r}}=\tilde{\mathbf{R}}_{\mathrm{CP}}\sum\nolimits_{p=1}^{P}\mathbf{G}_p\tilde{\mathbf{s}}+\mathbf{w}=\tilde{\mathbf{R}}_{\mathrm{CP}}\mathbf{G}\tilde{\mathbf{A}}_{\mathrm{CP}}\mathbf{s}+\mathbf{w}\nonumber\\
    &=\mathbf{G}_{\mathrm{T}}\mathbf{s}+\mathbf{w},\label{time domain IO}
\end{align}
where $\mathbf{G}_p$ represents the $p$-th resolvable path component of the time domain channel matrix $\mathbf{G}$ and the $(m,n)$-th element of $\mathbf{G}$ is given in \eqref{g_element}.
The OTFS CP removal matrix is defined as $\tilde{\mathbf{R}}_{\mathrm{CP}}\triangleq\left[\boldsymbol{0}_{MN\times L_{\mathrm{CP}}}, \mathbf{I}_{MN}\right]$ and $\mathbf{G}_{\mathrm{T}}\in\mathbb{C}^{MN\times MN}$ denotes the effective time domain channel matrix after OTFS CP removal.
Note that in \eqref{time domain IO} and what follows,
we reuse $\mathbf{w}$ for representing the noise vector since they follow the same distribution. Moreover, by further removing the OFDM CP from the receive signal, we obtain
\begin{align}
    \mathbf{r}'&=(\mathbf{I}_{\mathrm{N}}\otimes\mathbf{R}_{\mathrm{CP}})\mathbf{r}=(\mathbf{I}_{\mathrm{N}}\otimes\mathbf{R}_{\mathrm{CP}})\mathbf{G}_{\mathrm{T}}(\mathbf{I}_N\otimes\mathbf{A}_{\mathrm{CP}})\mathbf{s}'
\end{align}
where $\mathbf{R}_{\mathrm{CP}}\triangleq \left[\boldsymbol{0}_{M'\times L_{\mathrm{CP}}}, \mathbf{I}_{M'}\right]$ denotes the OFDM CP removal matrix.

Subsequently, the equivalent TF domain receive signals after OTFS and OFDM CP removal are then obtained by applying the DFT to the time domain signal $\mathbf{r}$ and $\mathbf{r}'$, respectively, yielding
\begin{align}
    \mathbf{y}_\mathrm{TF}
    &=\left(\mathbf{I}_N\otimes\mathbf{F}_{M}\right) \mathbf{G}_{\mathrm{T}}  \left(\mathbf{I}_N\otimes\mathbf{F}_{M}^{\mathrm{H}}\right)\mathbf{x}_\mathrm{TF}+\mathbf{w},\\
    &=\mathbf{H}_\mathrm{TF}\mathbf{x}_\mathrm{TF}+\mathbf{w},\label{IO_TF}\\ 
    \mathbf{y}'_\mathrm{TF}
    &=(\mathbf{I}_{\mathrm{N}}\otimes\mathbf{F}_{M'}\mathbf{R}_{\mathrm{CP}}) \mathbf{G}_{\mathrm{T}}  (\mathbf{I}_N\otimes\mathbf{A}_{\mathrm{CP}}\mathbf{F}_{M'}^{\rm H})\mathbf{x}'_\mathrm{TF}+\mathbf{w}, \\
    &=\mathbf{H}'_\mathrm{TF}\mathbf{x}'_\mathrm{TF}+\mathbf{w},\label{IO_TF_prime}
\end{align}
where $\mathbf{H}_\mathrm{TF}\in\mathbb{C}^{MN\times MN}$ and $\mathbf{H}'_\mathrm{TF}\in\mathbb{C}^{M'N\times M'N}$ represent the effective TF domain channel matrix after OTFS CP removal and OFDM CP removal, respectively. Considering a rectangular pulse shaping, $\mathbf{H}_\mathrm{TF}$ and $\mathbf{H}'_\mathrm{TF}$ can be expressed, respectively, as
\begin{align}
    \mathbf{H}_\mathrm{TF}&=\sum_{p=1}^{P} h_p e^{-j2\pi\frac{k_p l_p}{MN}} \left(\mathbf{I}_N\otimes\mathbf{F}_M\right) \boldsymbol{\Delta}^{k_p}   \boldsymbol{\Pi}^{l_p} \left(\mathbf{I}_N\otimes\mathbf{F}_M^{\mathrm{H}}\right)\label{H_TF}\\
     \mathbf{H}'_\mathrm{TF}&=(\mathbf{I}_{\mathrm{N}}\otimes\mathbf{F}_{M'}\mathbf{R}_{\mathrm{CP}}\mathbf{F}_{M}^{\mathrm{H}}) \mathbf{H}_\mathrm{TF}(\mathbf{I}_N\otimes\mathbf{F}_{M}\mathbf{A}_{\mathrm{CP}}\mathbf{F}_{M'}^{\rm H}).\label{H_prime}
\end{align}
More importantly, the relationship between these two TF domain transmitted symbol can be expressed as
\begin{align}
    \mathbf{x}_{\mathrm{TF}} = (\mathbf{I}_N\otimes\mathbf{F}_{M}\mathbf{A}_{\mathrm{CP}}\mathbf{F}_{M'}^{\rm H})  \mathbf{x}'_\mathrm{TF}. \label{x_prime}
\end{align}
Notice from \eqref{x_prime} that the $\mathbf{x}_{\mathrm{TF}}$ experience symbol spreading along the frequency dimension due to CP appending and removal, in contrast to $\mathbf{x}'_{\mathrm{TF}}$. This effect arises from the mismatch in DFT sizes, i.e., $\mathbf{F}_{M}$ versus $\mathbf{F}_{M'}$. As a result, interference between pilot and data symbols is inevitably introduced at the receiver $\mathbf{y}_{\mathrm{TF}}$. However, we perform channel estimation based on the model in \eqref{IO_TF} to obtain a structured channel representation that is suitable for subsequent cross-domain transformation.

\subsection{Properties of the TF Domain Effective Channel Matrix}
In this subsection, we exploit the structural properties of the TF domain effective channel matrix $\mathbf{H}_\mathrm{TF}$, which is utilized to facilitate the subsequent TF domain channel estimation. Specifically, following the conventional OFDM setup, let us partition $\mathbf{y}_\mathrm{TF}$ and $\mathbf{x}_\mathrm{TF}$ into $N$ symbol vectors, each containing $M$ samples (including the OFDM CP), such that $\mathbf{y}_\mathrm{TF}=[\mathbf{y}_{0}^\mathrm{T}, \mathbf{y}_{1}^\mathrm{T}, \dots, \mathbf{y}_{N-1}^\mathrm{T}]^\mathrm{T}$ and $\mathbf{x}_\mathrm{TF}=[\mathbf{x}_{0}^\mathrm{T}, \mathbf{x}_{1}^\mathrm{T}, \dots, \mathbf{x}_{N-1}^\mathrm{T}]^\mathrm{T}$. Moreover,  the channel matrix $\mathbf{H}_\mathrm{TF}$ exhibits a specific structure, which is described as follows

\begin{align}\label{H_TF_mtx}
\mathbf{H}_\mathrm{TF} = 
\begin{bmatrix}
\mathbf{H}_\mathrm{TF}^{0,0} & 0 & 0 & \cdots & 0 & \mathbf{H}_\mathrm{TF}^{0,1} \\
\mathbf{H}_\mathrm{TF}^{1,1} & \mathbf{H}_\mathrm{TF}^{1,0} & 0 & \cdots & 0 & 0 \\
0 & \mathbf{H}_\mathrm{TF}^{2,1} & \mathbf{H}_\mathrm{TF}^{2,0} & \cdots & 0 & 0 \\
\vdots & \vdots & \vdots & \ddots & \vdots & \vdots \\
0 & 0 & 0 & \cdots & \mathbf{H}_\mathrm{TF}^{N-2,0} & 0 \\
0 & 0 & 0 & \cdots & \mathbf{H}_\mathrm{TF}^{N-1,1} & \mathbf{H}_\mathrm{TF}^{N-1,0}
\end{bmatrix},
\end{align}
where $\mathbf{H}_\mathrm{TF}^{j,0}$ and $\mathbf{H}_\mathrm{TF}^{j,1}$, for $j=0,\dots,N-1$, denotes the $M\times M$ sub-channel matrices of $\mathbf{H}_\mathrm{TF}$. It is important to note that $\mathbf{H}_\mathrm{TF}^{j,0}$ is a dense matrix that captures the ICI induced by the Doppler effect, whereas $\mathbf{H}_\mathrm{TF}^{j,1}$ accounts for the ISI resulting from multipath delay. Based on the structure of $\mathbf{H}_\mathrm{TF}$, the IOR in (\ref{IO_TF}) can be expressed in block-wise as follows
\begin{align}
   {\mathbf{y}_\mathrm{TF}^{(i)}}=\mathbf{H}_\mathrm{TF}^{(i)}\mathbf{x}_\mathrm{TF}^{(i)} + \mathbf{w}^{(i)}, \label{IO_block}
\end{align}
where $\mathbf{x}_\mathrm{TF}^{(i)}\in\mathbb{C}^{M\times 1}, i=0,\dots,N-1$, denotes the $i$-th OFDM symbol that containing $M$ samples. Note that $\mathbf{H}_\mathrm{TF}^{(i)}\in\mathbb{C}^{2M\times M}$ and $\mathbf{y}_\mathrm{TF}^{(i)}\in\mathbb{C}^{2M\times 1}$ denote the corresponding channel matrix and observations, respectively, which can be expressed as 
\begin{align}
    \mathbf{H}_\mathrm{TF}^{(i)}=
    \begin{bmatrix}
\mathbf{H}_\mathrm{TF}^{i,0} \\
\mathbf{H}_\mathrm{TF}^{(i+1)_N,1} \\
\end{bmatrix},
    \mathbf{y}_\mathrm{TF}^{(i)}=
    \begin{bmatrix}
\mathbf{y}_\mathrm{TF}^{i,0} \\
\mathbf{y}_\mathrm{TF}^{(i+1)_N,1} \\
\end{bmatrix}.\label{y_i}
\end{align}
It is worth noting that the $i$-th OFDM symbol spreads into the subsequent symbol’s interval due to path delay, thereby resulting in a total observation length of $2M$ for $\mathbf{y}_\mathrm{TF}^{(i)}$. By following the same structure across all blocks, the overall received signal $\mathbf{y}_\mathrm{TF}$ can be constructed by aggregating these individual observations with appropriate zero-padding, as
\begin{align}
\mathbf{y}_\mathrm{TF} = 
\begin{bmatrix}
\mathbf{y}_\mathrm{TF}^{0,0}  \\
\mathbf{y}_\mathrm{TF}^{1,1} \\
0  \\
\vdots  \\
0  \\
0 
\end{bmatrix}
+
\begin{bmatrix}
 0  \\
\mathbf{y}_\mathrm{TF}^{1,0}  \\
 \mathbf{y}_\mathrm{TF}^{2,1}  \\
 \vdots  \\
 0  \\
 0 
\end{bmatrix}
+
\begin{bmatrix}
 0  \\
0  \\
 \mathbf{y}_\mathrm{TF}^{2,0}  \\
 \vdots  \\
 0  \\
 0 
\end{bmatrix}
+ \cdots+
\begin{bmatrix}
\mathbf{y}_\mathrm{TF}^{0,1} \\
 0 \\
 0 \\
 \vdots \\
 0 \\
 \mathbf{y}_\mathrm{TF}^{N-1,0}
\end{bmatrix}.\label{y_TF_mtx}
\end{align}

\begin{figure*}[ht]
    \centering
    \includegraphics[scale=0.46]{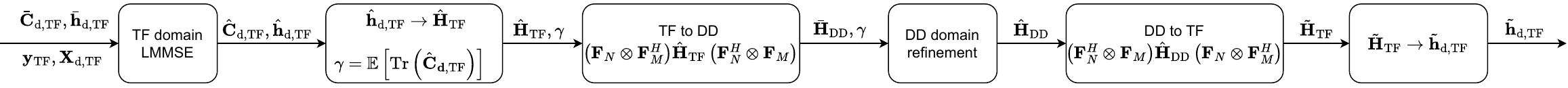}
    \caption{Block diagram for the CDCE scheme.}
    \label{fig:HCD}
\end{figure*}
\section{OFDM Channel Estimation via DD Domain Refinement}

\subsection{TF Domain Dense IOR}

Since the channel estimation only needs to focus on the non-zero elements of $\mathbf{H}_\mathrm{TF}$, we in the following derive a dense representation of the TF domain IOR by leveraging the structure of $\mathbf{H}_\mathrm{TF}$. Specifically, we begin by rewriting \eqref{IO_TF} as
\begin{align}
    \mathbf{y}_\mathrm{TF}
    &=\left(\mathbf{x}^{\mathrm{T}}_{\mathrm{TF}}\otimes\mathbf{I}_{MN}\right)\mathrm{vec}\left(\mathbf{H}_{\mathrm{TF}}\right)+\mathbf{w}=\mathbf{X}_{\mathrm{TF}}\mathbf{h}_{\mathrm{TF}}+\mathbf{w}.\label{IO_Xh}
\end{align}
Similarly, the block-wise IOR in (\ref{IO_block}) can be rewritten as
\begin{align}
   {\mathbf{y}_\mathrm{TF}^{(i)}}&=\left(\left(\mathbf{x}_\mathrm{TF}^{(i)}\right)^\mathrm{T}\otimes\mathbf{I}_{2M}\right)\mathrm{vec}\left(\mathbf{H}_\mathrm{TF}^{(i)}\right)+\mathbf{w}^{(i)}\notag\\
   &=\mathbf{X}^{(i)}_{\mathrm{TF}}\mathbf{h}^{(i)}_{\mathrm{TF}}+\mathbf{w}^{(i)},\label{IO_block_Xh}
\end{align}
where $\mathbf{h}^{(i)}_{\mathrm{TF}}\in\mathbb{C}^{2M^2\times 1}$ denotes the equivalent channel vector of the $i$-th OFDM symbol and $\mathbf{X}^{(i)}_{\mathrm{TF}}\in\mathbb{C}^{2M\times 2M^2}$ denote the corresponding equivalent transmit matrix. According to (\ref{IO_block})-(\ref{IO_block_Xh}), we can further rewrite (\ref{IO_Xh}) in a dense form as
\begin{align}
    \mathbf{y}_\mathrm{TF}&=\mathbf{X}_{\mathrm{d},\mathrm{TF}}\mathbf{h}_{\mathrm{d},\mathrm{TF}}+\mathbf{w}, \label{TF_dense_IO}
\end{align}
where $\mathbf{h}_{\mathrm{d},\mathrm{TF}}\in\mathbb{C}^{2M^2 N\times 1}$ is a dense format of $\mathbf{h}_{\mathrm{TF}}$ in (\ref{IO_Xh}), which is given by 
\begin{align}
    \mathbf{h}_{\mathrm{d},\mathrm{TF}}=\left[\left(\mathbf{h}^{(0)}_{\mathrm{TF}}\right)^{\mathrm{T}},\dots,\left(\mathbf{h}^{(i)}_{\mathrm{TF}}\right)^{\mathrm{T}},\dots,\left(\mathbf{h}^{(N-1)}_{\mathrm{TF}}\right)^{\mathrm{T}}\right]^{\mathrm{T}},
\end{align}
and the structure of $\mathbf{X}_{\mathrm{d},\mathrm{TF}}\in\mathbb{C}^{MN\times 2M^2 N}$ is given as

\begin{align}
\mathbf{X}_{\mathrm{d},\mathrm{TF}} = 
\begin{bmatrix}
\mathbf{X}_{\mathrm{TF}}^{0,0} & 0 & 0 & \cdots & 0 & \mathbf{X}_{\mathrm{TF}}^{0,1} \\
\mathbf{X}_{\mathrm{TF}}^{1,1} & \mathbf{X}_{\mathrm{TF}}^{1,0} & 0 & \cdots & 0 & 0 \\
0 & \mathbf{X}_{\mathrm{TF}}^{2,1} & \mathbf{X}_{\mathrm{TF}}^{2,0} & \cdots & 0 & 0 \\
\vdots & \vdots & \vdots & \ddots & \vdots & \vdots \\
0 & 0 & 0 & \cdots & \mathbf{X}_{\mathrm{TF}}^{N-2,0} & 0 \\
0 & 0 & 0 & \cdots & \mathbf{X}_{\mathrm{TF}}^{N-1,1} & \mathbf{X}_{\mathrm{TF}}^{N-1,0}
\end{bmatrix}.\label{X_d}
\end{align}
Note that $\mathbf{X}^{(i)}_{\mathrm{TF}}$ in (\ref{IO_block_Xh}) corresponds to a submatrix of $\mathbf{X}_{\mathrm{d}}$ in \eqref{X_d}, i.e.,
\begin{align}
    \mathbf{X}^{(i)}_{\mathrm{TF}}=\left(\left(\mathbf{x}_\mathrm{TF}^{(i)}\right)^{\mathrm{T}}\otimes\mathbf{I}_{2M}\right)=
    \begin{bmatrix}
\mathbf{X}_\mathrm{TF}^{i,0} \\
\mathbf{X}_\mathrm{TF}^{(i+1)_N,1} \\
\end{bmatrix}.
\end{align}

\subsection{TF Domain Estimation}
As illustrated in Fig.~\ref{fig:HCD}, the estimation process begins in the TF domain, where the inputs include the received signal $\mathbf{y}_{\mathrm{TF}}$, the equivalent pilot matrix $\mathbf{X}_{\mathrm{d},\mathrm{TF}}$, the \emph{a priori} mean of channel $\bar{\mathbf{h}}_{\mathrm{d},\mathrm{TF}}$, and its corresponding covariance matrix $\bar{\mathbf{C}}_{\mathrm{d},\mathrm{TF}}$. The parameters $\bar{\mathbf{h}}_{\mathrm{d},\mathrm{TF}}$ and $\bar{\mathbf{C}}_{\mathrm{d},\mathrm{TF}}$ are initialized through Monte Carlo simulations. Given these inputs, a coarse channel estimate is obtained through a full-size LMMSE estimator $\mathbf{V}$, whose expression is given as~\cite{chong2025cross}

\begin{align}
    \mathbf{V}=\bar{\mathbf{C}}_{\mathrm{d},\mathrm{TF}}\mathbf{X}_{\mathrm{d},\mathrm{TF}}^{\mathrm{H}}\left( \mathbf{X}_{\mathrm{d},\mathrm{TF}}\bar{\mathbf{C}}_{\mathrm{d},\mathrm{TF}} \mathbf{X}_{\mathrm{d},\mathrm{TF}}^{\mathrm{H}}+N_0\mathbf{I}_{MN} \right)^{-1}.\label{LMMSE}
\end{align}
The \emph{a posteriori} estimate $\hat{\mathbf{h}}_{\mathrm{d},\mathrm{TF}}$ and residual error covariance matrix $\hat{\mathbf{C}}_{\mathrm{d},\mathrm{TF}}$ are subsequently obtained by applying the LMMSE estimator as
\begin{align}
\hat{\mathbf{h}}_{\mathrm{d},\mathrm{TF}} &= \bar{\mathbf{h}}_{\mathrm{d},\mathrm{TF}}+\mathbf{V} \left(\mathbf{y}_{\mathrm{TF}}-\mathbf{X}_{\mathrm{d},\mathrm{TF}}\bar{\mathbf{h}}_{\mathrm{d},\mathrm{TF}}\right),\label{FS_LMMSE}\\
\hat{\mathbf{C}}_{\mathrm{d},\mathrm{TF}} &= \bar{\mathbf{C}}_{\mathrm{d},\mathrm{TF}}-\mathbf{V}\mathbf{X}_{\mathrm{d},\mathrm{TF}}\bar{\mathbf{C}}_{\mathrm{d},\mathrm{TF}}.\label{C_post}
\end{align}
It is worth noting that the trace of $\hat{\mathbf{C}}_{\mathrm{d},\mathrm{TF}}$ represents the mean square error (MSE) of the estimation when $\hat{\mathbf{h}}_{\mathrm{d},\mathrm{TF}}$ is an \emph{unbiased} estimate of the true channel, i.e., $\mathbb{E}[\hat{\mathbf{h}}_{\mathrm{d},\mathrm{TF}}]=\mathbf{h}_{\mathrm{d},\mathrm{TF}}$.

After performing LMMSE, the \emph{a posteriori} estimate $\hat{\mathbf{h}}_{\mathrm{d},\mathrm{TF}}$ is reshaped to matrix form $\hat{\mathbf{H}}_{\mathrm{TF}}$ with appropriate zero-padding, and subsequently transformed into the DD domain to serve as the initial channel estimate for further refinement. Specifically, the DD domain channel can be obtained by~\cite{liu2023predictive}
\begin{align}
    &\bar{\mathbf{H}}_{\mathrm{DD}}=\left(\mathbf{F}_N\otimes\mathbf{F}_M^{\mathrm{H}}\right)\hat{\mathbf{H}}_{\mathrm{TF}}\left(\mathbf{F}_N^{\mathrm{H}}\otimes\mathbf{F}_M\right),\label{TF to DD}\\
    &=\sum\nolimits_{p=1}^{P} h_p e^{-j2\pi\frac{k_p l_p}{MN}} \left(\mathbf{F}_N\otimes\mathbf{I}_M\right) \boldsymbol{\Delta}^{k_p}   \boldsymbol{\Pi}^{l_p} \left(\mathbf{F}_N\otimes\mathbf{I}_M^{\mathrm{H}}\right),\label{H_DD}
\end{align}
where $\left(\mathbf{F}_N\otimes\mathbf{F}_M^{\mathrm{H}}\right)$ denotes the symplectic finite Fourier transform (SFFT) and $\left(\mathbf{F}_N^{\mathrm{H}}\otimes\mathbf{F}_M\right)$ is its inverse operation.

\subsection{DD Domain Channel Matrix Properties}
To facilitate the DD domain channel estimation, let us first study the inherent structure of the DD domain channel matrix. Specifically, an \emph{arbitrary} DD domain channel matrix is characterized as follows
\begin{align}
{\mathbf{H}}_\mathrm{DD} =
\begin{bmatrix}
h_1 & 0 & h_P\phi(l,k) & \cdots  & 0 \\
0 & h_1\phi(l,k) & 0 & \cdots &  h_2\phi(l,k)\\
h_2 & 0 & h_1\phi(l,k) & \cdots &  0 \\
\vdots & \vdots & \vdots & \ddots  & \vdots \\
h_P & 0 & 0 & \cdots & 0 \\
0 & h_P\phi(l,k) & 0 & \cdots  & h_1\phi(l,k)
\end{bmatrix},\label{H_DD_structure}
\end{align}
where $h_i,1\le i\le P$, denotes the fading coefficient of the $i$-th path and $\phi(l,k)$ denotes the phase shifts introduced by the corresponding delay and Doppler indices. It can be noticed that each column of $\mathbf{H}_\mathrm{DD}$ corresponds to one observation of the DD domain discretized channel impulse response (with a specific location shift and phase shift). This implies that $\mathbf{H}_\mathrm{DD}$ contains $MN$ observations of the DD domain channel response. A useful interpretation for the location shifts across different observations is that the shifts within each observation are induced by a single DD domain pilot. The pilot positions vary across different observations and introduce corresponding shifts in the received signal through the twisted convolution. In other words, one can effectively envision $MN$ uniquely placed parallel pilots, each experiencing the DD domain channel, thereby yielding the corresponding observations in $\mathbf{H}_\mathrm{DD}$. For further illustration, let us consider two examples:
\begin{itemize}
    \item The first column of ${\mathbf{H}}_\mathrm{DD}$ corresponds to a pilot located at $ (l',k')=(0,0)$, yielding an observation that directly reflects a DD domain channel response without location shift.
    \item The second column of ${\mathbf{H}}_\mathrm{DD}$ corresponds to a pilot located at $(l',k')=(1,0)$, resulting in an observation of the DD domain channel response shifted by one unit along delay dimension and with corresponding phase shifts.
\end{itemize}
\noindent The above phenomenon arises from the twisted convolution property~\cite{nie2025novel} inherent to DD domain communications, which can be mathematically characterized (with noise omitted) as
\begin{align}
Y_{\mathrm{DD}}[l,k] 
&= {\alpha}_{p} 
X_{\mathrm{DD}}\left[\left[ l-l_p \right]_M,\left[ k - k_p \right]_N \right], \label{DD_IO_symbol_wise}
\end{align}
where $Y_{\mathrm{DD}} $ and $X_{\mathrm{DD}}$ denote the symbol-wise received and transmitted signal, respectively. Note that ${\alpha}_{p}$ corresponds to the non-zero elements in (\ref{H_DD_structure}), representing the symbol-wise effective channel coefficients that contain both the fading gain of the $p$-th path and the associated phase rotation, which is expressed as
\begin{equation}
{\alpha}_{p} =
\begin{cases}
 h_p \exp\left(j2\pi\frac{k_p(l-l_p)}{MN}\right), & l - l_p \geq 0, \\
 h_p \exp\left(j2\pi\frac{k_p(l-l_p)}{MN}\right)\exp\left(-j2\pi\frac{k - k_p}{N}\right), & l - l_p < 0 .
\end{cases}\label{alpha}
\end{equation}

\subsection{DD Domain Refinement}

Now, we are ready to present the DD domain channel estimation scheme. For a \emph{noisy} channel matrix $\bar{\mathbf{H}}_\mathrm{DD}$, the effective channel coefficient $\alpha_p$ can be estimated by applying an amplitude threshold, with values below the threshold suppressed to zero. To determine an appropriate threshold, the value $3\sqrt{N_0}$ from \cite{raviteja2019embedded} is no longer applicable due to the adoption of the LMMSE estimator in the TF domain. Instead, we define a modified threshold based on the average MSE of LMMSE estimation, which is computed from the \emph{a posteriori} error covariance matrix $\hat{\mathbf{C}}_{\mathrm{d},\mathrm{TF}}$ in (\ref{C_post}) by $\gamma = \frac{\mathrm{Tr}\left( \hat{\mathbf{C}}_{\mathrm{d},\mathrm{TF}} \right)}{2M^2N}$,
where $2M^2N$ is the total number of non-zero elements in the TF domain channel. 
Here, $\gamma$ represents the average estimation error of TF domain channel, which remains invariant under the unitary transformation from the TF domain to the DD domain. Therefore, our DD domain threshold is set to $3\sqrt{\gamma}$.  After obtaining the estimates of effective channel coefficients ${\alpha}_{p}$, the fading gain $h_{p}$ can be recovered by applying phase compensation to the estimates of ${\alpha}_{p}$.  Specifically, it can be observed from (\ref{DD_IO_symbol_wise}) and (\ref{alpha}) that the phase rotation explicitly depends on the indices $(l,k)$ in $Y_{\mathrm{DD}}$ and the indices of channel impulses $(l_{p},k_{p})$. Since $(l_{p},k_{p})$ are unknown, the objective is to infer the values of $(l_{p},k_{p})$ based on $(l,k)$ in $Y_{\mathrm{DD}}$ and $(l',k')=(\left[ l-l_p \right]_M,\left[ k - k_p \right]_N)$ in $X_{\mathrm{DD}}$. It is worth pointing out that $(l,k)$ in $Y_{\mathrm{DD}}$  can be inferred from the row indices of $\mathbf{H}_\mathrm{DD}$, while $(l',k')$ in $X_{\mathrm{DD}}$ can be inferred from the column indices of $\mathbf{H}_\mathrm{DD}$. By utilizing both sets of indices $(l,k)$ and $(l',k')$, the indices $(l_p,k_p)$ can be effectively determined, thereby enabling phase compensation according to (\ref{alpha}). After applying thresholding and phase compensation, $MN$ estimates of DD domain channel response are obtained. The next step is to average these $MN$ estimates and apply another thresholding operation, which can potentially reduce the probabilities of both missed estimations and false alarms.  The key steps of DD domain estimation are summarized in Algorithm~\ref{algo1}.
\begin{algorithm}
  \caption{DD domain refinement}
  \label{algo1}
  \begin{algorithmic}[1]
    \Statex \hspace{-\algorithmicindent}\textbf{Input:}
      $\bar{\mathbf{H}}_{\mathrm{DD}},\,
       \gamma$.
    \Statex \hspace{-\algorithmicindent}\textbf{Steps:}
    \State Set $\bar{\mathbf{H}}_{\mathrm{DD}}(i,j)=0, \forall |\bar{\mathbf{H}}_{\mathrm{DD}}(i,j)|<\gamma, 1\le i,j \le MN$.
    \State Compute $l'=[j-1]_M, k'=\lfloor \frac{j-1}{M} \rfloor,  l=[i-1]_M, k=\lfloor \frac{i-1}{M} \rfloor$.
    \State Compute $l_p=[l-l']_M, k_p=[k-k'+k_{\max}]_{N}-k_{\max}$.
    \State Compute phase by substituting $(l,k)$ and $(l_p,k_p)$ into (\ref{alpha}).
    \State Compensate phase, average estimates.
  \end{algorithmic}
\end{algorithm}
With the estimated channel responses obtained, the DD channel matrix can be reconstructed by substituting the estimated fading gains and corresponding delay and Doppler indices into (\ref{H_DD}). Subsequently, the estimated DD domain channel matrix $\hat{\mathbf{H}}_{\mathrm{DD}} $ is transformed back to the TF domain to yield the refined TF domain channel matrix as
\begin{align}
\tilde{\mathbf{H}}_{\mathrm{TF}} &= \left(\mathbf{F}_N^H \otimes \mathbf{F}_M\right) \hat{\mathbf{H}}_{\mathrm{DD}} \left(\mathbf{F}_N \otimes \mathbf{F}_M^H\right). \label{DD to TF}
\end{align}
Finally, the estimation of $\mathbf{H}'_{\mathrm{TF}}$ is obtained by substituting $\tilde{\mathbf{H}}_{\mathrm{TF}}$ into \eqref{H_prime}.

\section{Numerical Results}

In this section, we evaluate the performance of our proposed CE design through numerical simulations. Without loss of generality, an OFDM system with $M=N=16$ is considered. Each channel realization is modeled with $P = 3$ independently resolvable paths.  For each path, the delay and Doppler indices are uniformly drawn from $[0,l_{\max}]$ and $[-k_{\max},k_{\max}]$, respectively, where $l_{\max} = 2$ and $k_{\max} = 3$. Each channel coefficient is independently generated from a zero-mean complex Gaussian distribution with variance $\frac{1}{P}$. Moreover, a lattice-type TF pilot pattern is adopted~\cite{nie2025novel}. The TF domain pilots are placed in $\mathbf{x}'_{\mathrm{TF}}$ and set to all ones, while the data symbols are equally drawn from QPSK with unit power. The signal-to-noise ratio (SNR) is defined as $\frac{1}{N_0}$. The NMSE is defined as $\mathrm{NMSE}=\frac{||\hat{\mathbf{H}}-{\mathbf{H}}||^2}{||{\mathbf{H}}||^2}$.

To evaluate the performance of the proposed CE method, several conventional OFDM channel estimation schemes are considered as benchmarks. In particular, STE is applied at the pilot locations based on \eqref{IO_TF_prime} to estimate the channel state information (CSI), while the CSI at non-pilot locations is recovered through linear interpolation based on the nearest non-zero elements in both TF dimensions. The STE based on least square (LS) and LMMSE estimators is adopted, referred to as ``ST-LS'' and ``ST-LMMSE'', respectively. These estimators are expressed as
\begin{align}
   {\mathbf{h}'}_{\mathrm{ST-LS}}&=\mathbf{y}'_{\mathrm{TF}}\oslash\mathbf{x}'_{\mathrm{TF}},\\
   {\mathbf{h}'}_{\mathrm{ST-LMMSE}}&=\frac{1}{1+\frac{1}{\mathrm{SNR}}}\odot\hat{\mathbf{h}}_{\mathrm{ST-LS}},
\end{align}
where $\oslash$ and $\odot$ denote the element-wise division and Hadamard product, respectively. Note that the STE is performed under the assumption that the TF domain IOR is characterized as an element-wise product, which fails to account for the ICI effects during the estimation stage. Consequently, ${\mathbf{h}}_{\mathrm{ST-LS}}$ and ${\mathbf{h}}_{\mathrm{ST-LMMSE}}$ are of size $M'N \times 1$, representing only the diagonal components of ${\mathbf{H}}'_{\mathrm{TF}}$. For a fair comparison with proposed CE, we reshape the estimated channel vector as ${\mathbf{H}'}_{\mathrm{ST-LS}}=\mathrm{diag}\left(\hat{\mathbf{h}}_{\mathrm{ST-LS}}\right), 
    {\mathbf{H}'}_{\mathrm{ST-LMMSE}}=\mathrm{diag}\left(\hat{\mathbf{h}}_{\mathrm{ST-LMMSE}}\right)$, respectively.
These reconstructed matrices are then compared with the true TF domain channel matrix $\mathbf{H}'_{\mathrm{TF}}$ in terms of NMSE.  Another benchmark that captures the ICI effects is the full-sized LMMSE estimator, referred to as ``FS-LMMSE', which is implemented according to \eqref{FS_LMMSE}, where the final estimate ${\mathbf{H}'}_{\mathrm{FS-LMMSE}}$ is obtained through \eqref{H_prime}.

\begin{figure}
    \centering
    \includegraphics[scale=0.4]{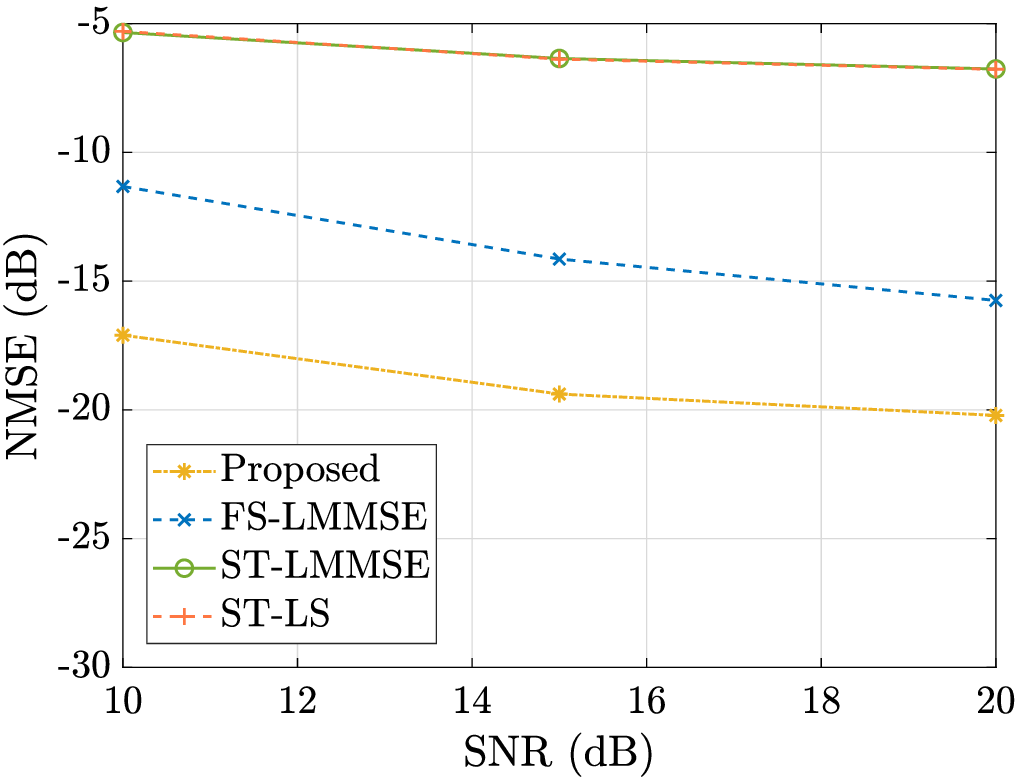}
    \caption{NMSE comparison between proposed CE and benchmarks with data symbols.}
    \label{fig:NMSE_compare_with_data}
\end{figure}

Fig.~\ref{fig:NMSE_compare_with_data} illustrates the NMSE performance comparison among the proposed CE method and all benchmark estimators. It is evident that the proposed CE consistently outperforms all other methods across various SNRs. In particular, while FS-LMMSE achieves moderate performance, the proposed CE achieves significantly lower NMSE due to DD domain refinement. In contrast, ST-LMMSE and ST-LS, which rely on STE followed by interpolation, exhibit saturated NMSE performance around $-6$ dB, as they fail to account for ICI. 

\section{Conclusion}

In this paper, we proposed a novel CE scheme for conventional OFDM systems by leveraging the advantages of DD domain channel. We began by exploiting the structure of the TF domain channel matrix to derive a dense IOR. Subsequently, a coarse TF domain channel estimate was obtained by adopting a full-sized LMMSE estimator, which was then transformed into the DD domain for further refinement. In the DD domain, the channel response was estimated through thresholding and phase compensation by exploiting the structure of the DD domain channel matrix. The resulting estimates obtained from different observations were subsequently averaged. Based on the estimated channel parameters, the DD domain channel matrix was reconstructed and transformed back to the TF domain to yield the final estimate.   Simulation results demonstrate that the proposed scheme consistently outperforms conventional OFDM estimators in terms of NMSE.

\bibliographystyle{IEEEtran}
\bibliography{ref}

@IEEEtranBSTCTL{bibfbcs:BSTcontrol,
	CTLuse_forced_etal = "yes",
	CTLmax_names_forced_etal = "1",
        CTLnames_show_etal       = "1"
}

@article{raviteja2019embedded,
    title={Embedded pilot-aided channel estimation for {OTFS} in delay-{Doppler} channels},
    author={Raviteja, Patchava and Phan, Khoa T and Hong, Yi},
    journal={IEEE Trans. Veh. Technol.},
    volume={68},
    number={5},
    pages={4906--4917},
    year={May 2019},
    }

@inproceedings{hadani2017orthogonal,
    title={Orthogonal time frequency space modulation},
    author={Hadani, Ronny and Rakib, Shlomo and Tsatsanis, Michail and Monk, Anton and Goldsmith, Andrea J and Molisch, Andreas F and Calderbank, R},
    booktitle={Proc. IEEE WCNC},
    pages={1--6},
    year={Mar. 2017},
}

@article{liu2023predictive,
  title={Predictive precoder design for {OTFS-enabled} {URLLC}: A deep learning approach},
  author={Liu, Chang and Li, Shuangyang and Yuan, Weijie and Liu, Xuemeng and Ng, Derrick Wing Kwan},
  journal={IEEE J. Sel. Areas Commun.},
  volume={41},
  number={7},
  pages={2245--2260},
  year={Jul. 2023},
}

@article{hu2024cross,
  title={Cross-Domain Channel Estimation Based Serial Interference Cancellation in {NOMA-OTFS} System},
  author={Hu, Jiacheng and Bai, Zhiquan and Xu, Hao and Liu, Hongwu and Wang, Yingxun and Kwak, KyungSup},
  journal={IEEE Commun. Lett.},
  volume={28},
  number={7},
  pages={1668--1672},
  year={Jul. 2024},
}

@article{chong2025cross,
  title={Cross-Domain Iterative Detection for {OTFS} Transmission with Frequency Domain Equalization},
  author={Chong, Ruoxi and Li, Shuangyang and Wei, Zhiqiang and Matthaiou, Michail and Ng, Derrick Wing Kwan and Caire, Giuseppe},
  journal={IEEE Trans. on Commun.},
  year={Jun. 2025},
  note={early access},
}

@article{nie2024uplink,
  title={Uplink multi-user {OTFS}: Transmitter design based on statistical channel information},
  author={Nie, Mingcheng and Li, Shuangyang and Mishra, Deepak and Yuan, Jinhong and Ng, Derrick Wing Kwan},
  journal={IEEE Trans. on Commun.},
  volume={73},
  number={7},
  pages={4678--4696},
  year={Dec. 2024},
}

@inproceedings{nie2025novel,
  title={A novel cross-domain channel estimation scheme for {OFDM}},
  author={Nie, Mingcheng and Chong, Ruoxi and Li, Shuangyang and Yuan, Weijie and Ng, Derrick Wing Kwan and Matthaiou, Michalis and Caire, Giuseppe and Li, Yonghui},
  booktitle={IEEE GLOBECOM Proceedings},
  year={2025},
}

@article{liu2024fast,
  title={Fast computation of zero-forcing precoding for massive {MIMO-OFDM} systems},
  author={Liu, Junkai and Zhang, Wei and Jiang, Yi},
  journal={IEEE Trans. Signal Process.},
  volume={72},
  pages={912--927},
  year={Jan. 2024},
}

@ARTICLE{11373535,
  author={Nie, Mingcheng and Chong, Ruoxi and Li, Shuangyang and Farhang, Arman and Göttsch, Fabian and Ng, Derrick Wing Kwan and Matthaiou, Michail and Li, Yonghui},
  journal={IEEE Commun. Stand. Mag.}, 
  title={Toward Standardizing {OTFS}: A Candidate Waveform for Next-Generation Wireless Networks}, 
  year={Feb. 2026},
  volume={},
  number={},
  pages={1-12},
  doi={10.1109/MCOMSTD.2026.3657606}}

@inproceedings{nie2023improving,
  title={Improving channel estimation performance for uplink {OTFS} transmissions: Pilot design based on a posteriori {Cram{\'e}r-Rao} bound},
  author={Nie, Mingcheng and Li, Shuangyang and Mishra, Deepak},
  booktitle={Proc. IEEE Int. Conf. Commun. Workshops (ICC Workshops)},
  pages={301--306},
  year={2023},
}

@ARTICLE{xiaoqi2025otfs,
  author={Zhang, Xiaoqi and Ni, Zhitong and Yuan, Weijie and Andrew Zhang, J. and Quek, Tony Q. S.},
  journal={IEEE Trans. Commun.}, 
  title={Deep Learning-based OTFS Channel Estimation and Symbol Detection with Plug-and-Play Framework}, 
  year={2025},
  volume={},
  number={},
  pages={1-1},}

@ARTICLE{xiaoqiprior2025,
  author={Zhang, Xiaoqi and Liu, Chang and Yuan, Weijie and Zhang, J. Andrew and Ng, Derrick Wing Kwan},
  journal={IEEE Trans. Vehicular Tech.}, 
  title={Sparse Prior-Guided Deep Learning for OTFS Channel Estimation}, 
  year={2024},
  volume={73},
  number={12},
  pages={19913-19918}}

\newpage

\vfill

\end{document}